\documentclass[aps, prx, amssymb, longbibliography, twocolumn, superscriptaddress, 10pt]{revtex4-2}

\usepackage[ruled,lined]{algorithm2e}
\usepackage{algpseudocode} 
\usepackage[german,english]{babel}
\usepackage[utf8]{inputenc}
\usepackage{makecell}
\usepackage{booktabs}
\usepackage{graphicx}
\usepackage{dcolumn}
\usepackage{bm}
\usepackage{braket}
\usepackage{amsmath}
\usepackage{tabularx}
\usepackage[normalem]{ulem}
\usepackage{listings}
\usepackage{xcolor}
\usepackage{tikz}
\usepackage{hyperref}
\hypersetup{                    
	colorlinks,
	citecolor=blue,
	filecolor=blue,
	linkcolor=blue,
	urlcolor=blue
}

\begin{document}

\title{Efficient classical simulation of time dynamics in Fermi-Hubbard models with imaginary interactions}
	
\author{Raul A. Santos}
\affiliation{Phasecraft Ltd.}

\date{\today}
	
\begin{abstract}
    Using a map between the Lindbladian evolution of dephasing in free fermions and the time evolution of imaginary-interaction Fermi-Hubbard models in bipartite lattices, we present an efficient classical algorithm to solve the Schr\"{o}dinger equation in these interacting systems. This algorithm leverages the recently discovered algorithm for simulating Lindbladian evolution by sampling mixed unitary channels \cite{wang2026}. 
    We comment on the expected classical complexity of the problem for general complex values of the parameters and discuss some applications.
\end{abstract}

\maketitle

Interactions in fermionic systems are fundamental to generate the infinitely rich phenomena observed in the world. Understanding their role in the behaviour of many-particle systems has been an incredibly challenging and rewarding endeavour in the natural sciences. The difficulty of finding a classical approach to capture them efficiently in general has motivated the study of minimal models where the role of interactions can be studied in detail. Among these models, a paradigmatic example is the Fermi-Hubbard model \cite{Hubbard1963, Kanamori_1963}. Despite its simplicity, its phase diagram has not been fully understood yet \cite{Arovas_2022, Qin_2022}. In 1D, the model is exactly solvable through the Bethe ansatz technique \cite{Lieb_1968,Essler2005}. In higher dimensions, several numerical and theoretical techniques have been developed to understand its low temperature behavior, ranging from exact diagonalisation, tensor networks and quantum Monte Carlo methods, each with their own range of applicability \cite{LeBlanc_2015}.
The non-equilibrium scenario is more challenging. Time dynamics simulation is known to be classically harder due to the expected growth on entanglement \cite{Ho_2017}. This difficulty has been particularly showcased by recent experiments on quantum computers, \cite{phasecraft_quantinuum2025,phasecraft_google2025,granet2025}. Moreover, it has been shown that dynamics of wavepackets in the Hubbard model can reproduce universal quantum computation \cite{Childs_2013, Bao_2015}, providing another evidence on the intrinsic difficulty of classically modeling this system in general.

A natural extension of the Fermi-Hubbard model is to allow arbitrary complex parameters in the Hamiltonian \cite{Ashida_2020}. Although this makes the Hamiltonian non-Hermitian, it still defines an operator with a minimum energy as long as the Hamiltonian posses a  $\mathcal{PT}$ symmetry \cite{Bender_1998}. These extensions capture the effects of dissipation in a Lindblad approach \cite{Torres_2014,takemori2024} and lead to several phenomena ranging from a significant alteration of the excitations above the Hubbard gap and appearance of novel exceptional points \cite{Nakagawa_2021, Heiss_2012} to dynamical instabilities of the anti-ferromagnetic Mott phase \cite{Pan_2020}.

In this letter we present a classically efficient algorithm to extract the time evolved many-body wavefunction of Fermi-Hubbard models with imaginary interactions (see Eq. \ref{eq:HUb_imag}). We achieve this by leveraging the map from the Lindbladian evolution of free fermions with dephasing and time evolution of imaginary Fermi-Hubbard models. Classically simulating Lindbladian dynamics is a formidable task as it entails evolving the full density matrix instead of a state. Recently Wang et al. \cite{wang2026} have introduced a new algorithm to simulate Lindbladian dynamics of dephasing by sampling unitary channels. Using this algorithm, the simulation of dephasing in non-interacting fermion systems becomes classically efficient, and through the mapping with imaginary Fermi-Hubbard evolution it provides an efficient classical approach to simulate those systems. Extending the discussion to the simulation of generalized Fermi-Hubbard models where the parameters can be any complex number, we comment on the classical complexity landscape, that takes the shape of a solid torus.

{\it Dephasing and imaginary Fermi-Hubbard models.-} 
We consider the open dynamics generated by the Lindbladian
\begin{align}\label{eq:Lindblad_deph}
    \mathcal{L}(\rho)=-i[H,\rho]+\gamma\sum_i \left (n_i\rho n_i -\frac{1}{2}\{n_i,\rho\}\right),
\end{align}
with $H:=\sum_{i,j\in\mathcal{G}}h_{ij}c_i^\dagger c_j$ and $n_i:=c_i^\dagger c_i$ and $\gamma\geq 0$. The operators $c_i$ $(c_i^\dagger)$ destroy (create) fermions at site $i$ in the bipartite lattice $\mathcal{G}=\mathcal{A}\cup\mathcal{B}$ and satisfy the canonical anticommutation relations $\{c_i,c_j^\dagger\}=\delta_{ij}$. We assume that $h_{ij}$ is a symmetric matrix with nonzero real entries only for terms connecting between the $\mathcal{A}$ and $\mathcal{B}$ sublattices.

The time evolved density matrix $\rho(t)=e^{t\mathcal{L}}(\rho_0)$ of a system with $M$ particles can be written as
\begin{align}
\rho(t):=\sum_{\bm{n,m}}\Psi_{\bm{nm}}(t)\prod_{i=1}^{M}c_{n_{i}}^{\dagger}|0\rangle\langle0|\prod_{j=1}^{M}c_{m_{j}},
\end{align}
using the fermionic algebra and the properties of the trace, the wavefunction
 $\Psi_{\bm{nm}}(t)$ is explicitly given by
\begin{align}
    {\rm Tr}\left(\prod_{j=1}^{M}c_{m_{j}}^{\dagger}\prod_{i=1}^{M}c_{n_{j}}\rho(t)\right)=\Psi_{\bm{nm}}(t).
\end{align}

Vectorising the density matrix it is possible to express this generator of time dynamics in the thermofield representation \cite{Umezawa1995,Medvedyeva_2016}
\begin{align}\label{eq:Lind_vec}
    \mathcal{L}=-i(H-\bar{H})+\gamma\sum_j \left( n_j\bar{n}_j-\frac{1}{2}(n_j+\bar{n}_j)\right)
\end{align}
where $\bar{H}=\sum_{l,j}h_{ij}\bar{c}_l^\dagger\bar{c}_j$. We have introduced the auxiliary fermionic operators $\bar{c}_i$ that act on the vectorized density matrix 
\begin{align}
|\rho(t)\rangle:=\sum_{\bm{n,m}}\Psi_{\bm{nm}}(t)\prod_{j=1}^{M}\bar{c}^\dagger_{m_{j}}\prod_{i=1}^{M}c_{n_{i}}^{\dagger}|0\rangle\otimes|0 \rangle
\end{align}
by left multiplication \cite{Umezawa1995}. The thermofield representation of the Lindbladian Eq. \ref{eq:Lind_vec} has a $\mathcal{PT}$ symmetry generated by $\mathcal{P}c_j\mathcal P^\dagger=\bar{c}_j$ and $\mathcal{T}i\mathcal{T}=-i$. This ensures that the spectrum of $\mathcal{L}$ has a $D_2$ symmetry \cite{Prosen_2012}, where the eigenvalues are real or come in complex conjugate pairs \cite{Bender_2003}. The form of $-i\mathcal{L}$ strongly resembles a generalized Fermi-Hubbard model, where the interaction is occurring between the bar and unbar fermionic modes. This connection can be sharpened by using the sublattice property of $h_{ij}$ and the unitary operators
\begin{align}
    \mathcal{U}_\mathcal{A}(\theta):=e^{i\sum_{j\in\mathcal{A}} \theta \bar{n}_j},\quad \mathcal{U}_\mathcal{B}(\theta):=e^{i\sum_{j\in\mathcal{B}} \theta \bar{n}_j}.
\end{align}
Note that conjugating the fermion operators in the sublattice $\mathcal{A}(\mathcal{B})$ by the unitary transformation $\mathcal{U}_{\mathcal{A}(\mathcal{B})}(\pi)$ they acquire a minus sign. Unitarily transforming $\mathcal{L}$ with $\mathcal{U}_A(\pi)$ and relabelling the modes $c_i:= a_{i,\downarrow}, \bar{c}_i:=a_{i,\uparrow}$ and $n_{i,\sigma}:=a_{i,\sigma}^\dagger a_{i,\sigma}$ the explicit connection with Fermi-Hubbard models becomes $\mathcal {U}_{\mathcal{A}}(\pi)(i\mathcal{L}) \mathcal {U}^\dagger_{\mathcal{A}}(\pi):=\mathcal{H}$ with
\begin{align}\label{eq:HUb_imag}
    \mathcal{H}=\sum_{ij,\sigma}h_{ij}a^\dagger_{i,\sigma}a_{j,\sigma}+i\gamma \sum_i n_{i\uparrow} n_{i\downarrow}-i\frac{\gamma}{2}\sum_{i,\sigma}n_{i,\sigma}
\end{align}
This model has several symmetries \cite{Heilmann1971,Yang1989,Yang1990,Essler2005}. $U(1)_c$ total charge symmetry is generated by the operator $Q:=\sum_i(n_{i,\uparrow}+n_{i,\downarrow})$. The spin operators $
    S_\alpha=\frac{1}{2}\sum_{i,\sigma,\sigma'}a^\dagger_{i,\sigma}(\sigma_\alpha)_{\sigma\sigma'}a_{i,\sigma'}$
where $\sigma_\alpha=(\sigma_x,\sigma_y,\sigma_z)$ is any of the three Pauli matrices commute with $\mathcal{H}$ and are a representation of an $SU(2)$ algebra. Unitary transformations generated by these operators correspond to rotations of the spin direction.
At half filling, another symmetry is present generated by the operators
$\eta^z=\sum_{i}(n_{i,\uparrow}+n_{i,\downarrow}-1)$, $\eta^+:=\sum_{j\in\mathcal{A}}c_{j\uparrow}^\dagger c_{j,\downarrow}^\dagger-\sum_{j\in\mathcal{B}}c_{j\uparrow}^\dagger c_{j,\downarrow}^\dagger$, and $\eta^-=(\eta^+)^\dagger$. These operators generate a second $SU(2)$ symmetry.

The connection between Fermi-Hubbard models with imaginary interactions and the evolution of dephasing spinless fermions presented above provides an alternative route to simulating the interacting system. Making use of the recently developed algorithm \cite{wang2026} for approximating dephasing evolution by uniformly sampling unitaries (which we explain below), the simulation of the open quantum system becomes classically efficient as it just corresponds to sampling unitaries generated by quadratic fermion operators.

{\it Classical simulation of dephasing models.-}
The time evolution with the unital Lindblad operator 
\begin{align}
    \mathcal{L}(\rho)=-i[H,\rho]+\sum_{i=1}^N \left( \ell_i \rho \ell_i - \frac{1}{2}\{\ell_i^2,\rho\}\right),
\end{align}
where $\ell_i$ is hermitian can be approximated by the expectation value of the stochastic channel 
\begin{align}\label{eq:stoch_channel}
   \mathcal{U}_{t,\bm s}(\rho):= e^{-iHt} \left(\prod_{j=1}^N e^{s_ji\sqrt{t}\ell_j}\rho e^{-s_ji\sqrt{t}\ell_j}\right)e^{iHt},
\end{align}
where each $s_j=\pm 1$ in ${\bm s}=(s_1,\dots s_N)$ is a uniformly distributed random variable as recently shown in \cite{wang2026}. The reason of this can be understood by considering the expected channel after a single Trotter step,
\begin{align}\nonumber
    \mathcal{E}(\rho):&=e^{-iHt}\frac{1}{2}(e^{i{\sqrt{t}}L}\rho e^{-i{\sqrt{t}}L}+e^{-i{\sqrt{t}}L}\rho e^{i\sqrt{t}{L}})e^{iHt}\\&=e^{t\mathcal{L}}(\rho)+O(t^{2}),
\end{align}
where we have included a single jump operator for simplicity. Expanding the channel to first order in time one finds $
    \mathcal{L}=-i[H,\rho]+L\rho L -\frac{1}{2}\{L^2,\rho\}.$
As shown in \cite{wang2026} the error incurred in this approximation for arbitrary time is
\begin{align}
\left\|\mathbb{E}(\mathcal{U}^R_{t/R,\bm s}(\rho))-e^{t\mathcal{L}}(\rho)\right\|_\diamond= O(t^2/R),
\end{align}
where $R$ is the number of Trotter steps. Setting $\ell_i=n_i$, we recover the dephasing Lindbladian Eq. \ref{eq:Lindblad_deph}. Here $\mathcal{U}^R_{t/R,\bm s}(\rho)$ is the stochastic map in Eq. \ref{eq:stoch_channel} iterated $R$ times, where each application is a new random instance.
Combining this method to efficiently simulate Lindbladians withe the map between Lindbladian evolution and time dynamics of Fermi-Hubbard models provides a route to efficiently simulate these systems by the duality
\begin{align}
     e^{-it\mathcal{H}}|\psi_\uparrow\rangle|\phi_\downarrow\rangle \leftrightarrow \mathcal {U}_{\mathcal{A}}(\pi)\left(e^{t\mathcal{L}}(|\psi\rangle\langle\phi|)\right)\mathcal{U}^\dagger_{\mathcal{A}}(\pi)
\end{align}

The algorithm to simulate the right hand side above is presented in Alg. \ref{alg:class_sim}.

\begin{algorithm}
    \caption{Classical simulation of imaginary Fermi-Hubbard models}
    \label{alg:class_sim}
    \begin{algorithmic}[1]
        \Procedure{Wavefunction}{$\bm{n},\bm{m},t,R, n_\uparrow,n_\downarrow, M, \Psi_0$} \Comment{An approximation to the time evolved wavefunction $\Psi_{\bm{nm}}(t)$ with $R$ steps, over time $t$ and $n_\sigma$ particles of spin $\sigma$ using $M$ samples}
            \State $\rho_0 \gets |\psi_\uparrow\rangle\langle \psi_\downarrow|$ \Comment{The initial state is assumed to be of the form $|\psi_\uparrow\rangle|\psi_\downarrow\rangle$}
            \State \For{$i=1\dots M$}{
            Sample $\bm{s}=((s_{jk})_{k=1}^N)_{j=1}^R$ where $s_{jk}=\{-1,1\}$ is sampled from the uniform distribution.\\
            $\rho_i(t) \gets \mathcal{U}^R_{t/R,\bm s}(\rho_0)$\\
            $\Psi^{(i)}_{\bm{nm}}(t)\gets {\rm Tr}\left(\prod_{j=1}^{n_\downarrow}c_{m_{j}}^{\dagger}\prod_{p=1}^{n_\uparrow}c_{n_{p}}\rho_m(t)\right)$}
            \State \textbf{return} $\frac{1}{M}\sum_m \Psi^{(m)}_{\bm{nm}}(t)$\Comment{This is the sample mean estimate of $\Psi_{\bm{nm}}(t)$}
        \EndProcedure
    \end{algorithmic}
\end{algorithm}

The sample complexity of this approach can be determined by the standard Hoeffding's inequality \cite{Hoeffding_1963}. Let $X\in [x_{min},x_{min}+\Delta]$ be a random variable, then the probability that the sample mean of $M$ observations 
$\hat{x}_M:=\frac{1}{M}\sum_{i=1}^MX_i$ deviates from the true mean $\bar{x}$ by $\epsilon$ is
    $\mathbb{P}(|\hat{x}_n-\bar{x}|> \epsilon )=2\exp\left(-\frac{2M\epsilon^2}{\Delta^2}\right)$.
The number of measurements needed to achieve a confidence $1-\delta$ that the sample mean is within $\epsilon$ from the true mean is then lower bounded by
\begin{align}
    M\geq \frac{\Delta^2}{2\epsilon^2}\log\left(\frac{2}{\delta}\right).
\end{align}
The crucial parameter that determines the required number of samples and the efficiency of the algorithm is the range that the random variable can take. Considering the expectation value of a time-evolved bounded operator $A$, 
\begin{align}\nonumber
    \langle A(t)\rangle&:= {\rm Tr}(e^{t\mathcal{L}}(\rho)A)\\&=\sum_{\bm s} p_{\bm s} {\rm Tr}(\mathcal{U}^R_{t/R,\bm s}(\rho)A)+O(t^2/R),
\end{align}
the sample complexity of approximating this sum by sampling the random variable $X(\bm s)={\rm Tr}(\mathcal{U}^R_{t/R,\bm s}(\rho)A)$ can obtained by bounding the  the range of $X(\bm s)$ using H\"{o}lder's inequality
\begin{align}
    |X(\bm s)|=|{\rm Tr}(\mathcal{U}^R_{t/R,\bm s}(\rho)A)|\leq \|A\|_\infty
\end{align}
where $\|A\|_\infty$ is the operator norm of $A$. Here have used the unitarity of the channel $\mathcal{U}_{t/R,{\bm s}}$ and that $\rho$ is a density matrix.

{\it Numerical results.-}
To showcase our results in practice we simulate the exact Hamiltonian evolution of a 2-site Fermi-Hubbard model described by the Hamiltonian
\begin{align}\nonumber
    H &= -J\sum_{\sigma=\uparrow,\downarrow}(c^\dagger_{1,\sigma}c_{2,\sigma}+c^\dagger_{2,\sigma}c_{1,\sigma})+iU\sum_{j=1,2}n_{j,\uparrow}n_{j,\downarrow}\\
    &-i\frac{U}{2}\sum_{j=1,2}(n_{j\uparrow}+n_{j\downarrow})
\end{align}
with initial state $|\Psi_0\rangle=c^\dagger_{1,\uparrow}c^\dagger_{2,\downarrow}|0\rangle$.
Here we fix $U=J=1$ and evolve for up to time $t=10$.
\begin{figure}[h!]
    \centering
    \includegraphics[width=\linewidth]{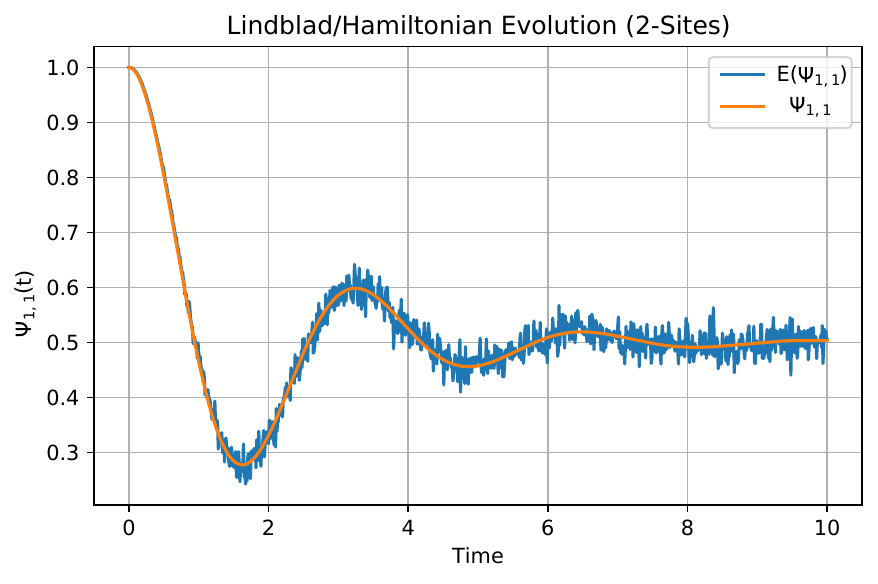}
    \caption{2-particle wavefunction $\Psi_{1,1}$ with $n_\uparrow=n_\downarrow=1$. Here $\Psi_{1,1}$ is computed by direct solution of the (complexified) Schr\"{o}dinger equation for $J=1$, $U=1$. $\mathbb{E}(\Psi_{1,1})$ is the sample mean obtained with the classical protocol Alg. \ref{alg:class_sim}. This results are obtained with 200 samples per data point, $t/R=0.01$ and $R=1000$.}
    \label{fig:numerics}
\end{figure}
We solve directly by matrix exponentiation the Schr\"{o}dinger equation $|\Psi(t)\rangle=e^{-itH}|\Psi_0\rangle$ and compare value of the wavefunction $\Psi_{1,1}(t)$ with the result obtained from
Alg \ref{alg:class_sim}, where we evolve with a spinless fermionic Hamiltonian
\begin{align}
    H_{dual}= -J(c^\dagger_{1}c_{2}+c^\dagger_{2}c_{1})
\end{align}
and an exponential of the density operators with random times $\pm\sqrt{tU}$ sampled uniformly. The results are shown in Fig. \ref{fig:numerics}

{\it Generalisations and complexity landscape.-} As we have discussed it is possible to classically approximate the time evolution of Fermi-Hubbard models with imaginary interaction. A natural question is how this relates with the standard time evolution corresponding to real interactions. We can capture the time evolution for any complex parameter by the following definition of time evolved state
\begin{align}
    |\rho(t,J,U)\rangle:&=e^{-it(JH_{0}+UH_{1})}|\rho_0\rangle \\\nonumber
    &=\sum_{\bm{n,m}}\Psi_{\bm{nm}}(t,J,U)\prod_{j=1}^{M}\bar{c}^\dagger_{m_{j}}\prod_{i=1}^{M}c_{n_{i}}^{\dagger}|0\rangle\otimes|0 \rangle
\end{align}
where $J,U \in \mathbb{C}$ and $t\in \mathbb{R}$. Here $H_0$ is a quadratic fermionic Hamiltonian corresponding to the free fermion evolution. $H_1:=\sum_in_{i\uparrow}n_{i,\downarrow}$ describes the fermion interactions. The usual Hamiltonian time evolution corresponds to the slice $J,U\in \mathbb{R}$. 
As before, the wavefunction $\Psi_{\bm n, \bm m}(t,J,U)$
can be obtained by evolving with the complexified Lindbladian
\begin{align}
    \Psi_{\bm n, \bm m}(t,J,U)&={\rm Tr}(e^{t\mathcal{L}(J,U)}(\rho_0)\mathcal{O}_{\bm n, \bm m})
\end{align}
where $\mathcal{O}_{\bm n, \bm m}=\prod_{j=1}^{M}c_{m_{j}}^{\dagger}\prod_{i=1}^{M}c_{n_{j}}$ and $\mathcal{L}(J,U):=-iJ[H_0,\cdot]+U\sum_i( n_i (\cdot)n_i-\frac{1}{2}\{n_i,\cdot\})$.

\begin{figure}[h!]
    \centering
    \includegraphics[width=\linewidth]{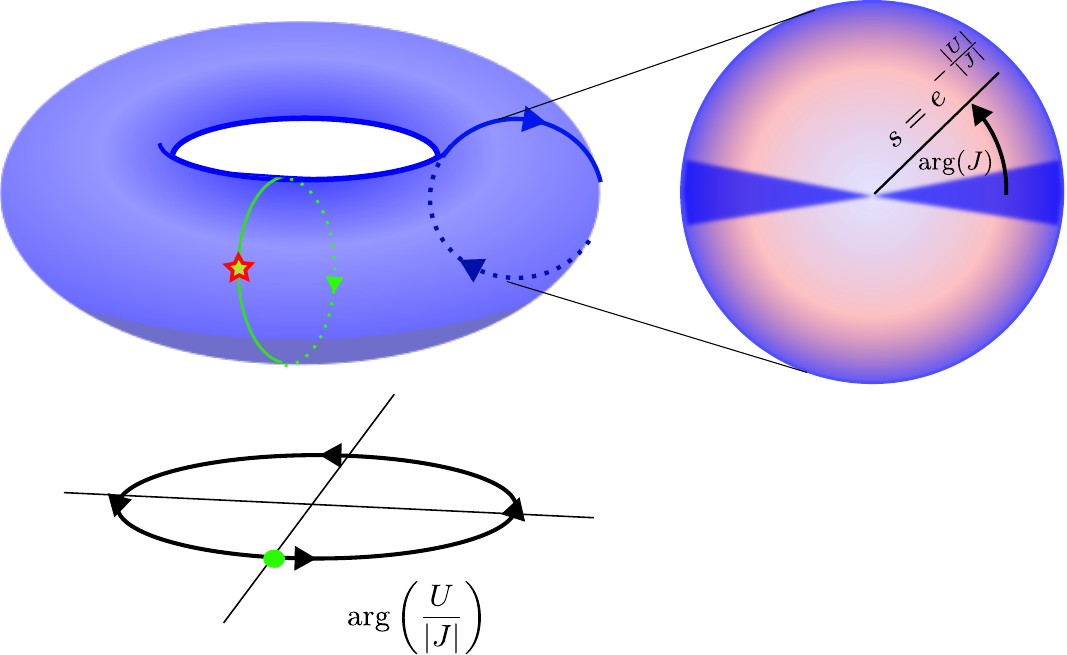}
    \caption{Classical complexity landscape of time evolution in (complexified) Fermi-Hubbard models (Complexity doughnut). The yellow star ${\rm arg}(J)={\rm arg}(U)=0$ represents the time evolution of usual Fermi-Hubbard models. Time dynamics simulation of these is believed to be difficult classically for generic values of the parameters. The surface of the torus $s=1\rightarrow U=0$ on the contrary is efficiently simulable due to the algebraic structure of free fermion Hamiltonians.  The results of this work shows that on the disk ${\rm arg}(U)=\frac{\pi}{2}$ the line ${\rm arg}(J)=0$ is also classically simulable. The classically accessible region with the method of this work extends over $\Im(J)=O((Lt)^{-1})$ where $L$ is the size of the system and $t$ is the simulation time.}
    \label{fig:landscape}
\end{figure}

We can characterize the manifold of parameters that defines the many-body wavefunction as a solid torus by using the parameterisation
\begin{align}\label{eq:parametrisation}
    \Psi_{\bm n, \bm m}(t,J,U)=\Psi_{\bm n, \bm m}(t|J|,e^{i{\rm arg}(J)},\frac{|U|}{|J|}e^{i{\rm arg}(U)}),
\end{align}
introducing $s:=e^{-\frac{|U|}{|J|}}$ the manifold $\mathcal{M}=S^{1}\times S^{1}\times[0,1]$ associated with $({\rm arg}(J),{\rm arg}(U), s)$ fully determines the state for a fixed evolution time $t|J|$. A sketch of this manifold is shown in Fig. \ref{fig:landscape}.

To quantify the classical complexity of computing $|\Psi(t,J,U)\rangle$ we are implicitly assuming that the initial state is classically computable, and in particular that corresponds to a tensor product of fermionic gaussian states in each spin sector. 
Some regions of the complexity doughnut are known. With the parameterisation of Eq. \ref{eq:parametrisation}, the surface of the doughnut $(s=1\rightarrow U=0)$ is classically solvable. This is due to the algebraic structure of quadratic Hamiltonians that map fermion operator into linear combinations of fermion operators. This structure does not rely on the unitarity of the operators, but instead in the quadratic nature of the generators $H_0$ of similarity transformations $e^{itJH_0}$ for $J\in \mathbb{C}$. 
On the other hand the time evolution of the usual Fermi-Hubbard Hamiltonian ${\rm arg}(J)={\rm arg}(U)=0$ is expected to be difficult classically \cite{Bao_2015}. 

The method that we discuss in this work becomes inefficient for generic points in the complexity doughnut. One way of seeing this is considering again the sample complexity of the Monte Carlo sum 
\begin{align}
    \Psi_{\bm n, \bm m}(t,J,U)&={\rm Tr}(e^{t\mathcal{L}(J,U)}(\rho_0)\mathcal{O}_{\bm n, \bm m})\\\nonumber
    &=\sum_{\bm s}p_{\bm s}{\rm Tr}(S^R_{t/R,{\bm s}}\rho_0 S^{-1}_{t/R,\bm s}\mathcal{O}_{\bm n,\bm n})+O(t^2/R)
\end{align}
where $S$ is now a similarity transformation 
\begin{align}
    S_{t,\bm s}(\rho):= e^{-iJH_0t}\prod_{j=1}^N e^{s_ji\sqrt{tU}n_j}
\end{align}
for $J,U\neq \mathbb{R}$, instead of a unitary. This changes the range of the random variable $\tilde{X}(\bm s)={\rm Tr}(S^R_{t/R,{\bm s}}\rho_0 S^{-1}_{t/R,\bm s}\mathcal{O}_{\bm n,\bm n})$ because the operator norm is not invariant under similarity transformations, meaning that in this case the variance of $X$ could depend exponentially on the system size. We can see this directly fixing $U\in \mathbb{R}$ but letting $J \in \mathbb{C}$. Using H\"{o}lder's inequality again we find
\begin{align}\nonumber
    |\tilde{X}(\bm s)|&\leq \|(S^{-1}_{t/R,\bm s})^R\mathcal{O}_{\bm n,\bm n}S^R_{t/R,{\bm s}}\|_\infty\\&\leq \|S^{-1}_{t/R,\bm s}\|_\infty\|\mathcal{O}_{\bm n,\bm n}\|_\infty \|S^R_{t/R,{\bm s}}\|_\infty
\end{align}
where we have used the sub-multiplicativity of the operator norm. As $S$ is generated by a quadratic fermion Hamiltonian, it is easy to compute its operator norm
\begin{align}\nonumber
    \|S^R_{t/R,{\bm s}}\|_\infty&\leq \|e^{-itJH_0}\|_\infty= \|Ue^{itJ\sum_j \varepsilon_jn_j}U^\dagger\|_\infty\\&=\exp\left(-t\Im(J)\min_{n_j\in\{0,1\}}\sum_j \varepsilon_j n_j\right)
\end{align}
where we have assumed that $\Im(J)<0$ (for $\Im(J)>0$ is the norm of $S^{-1}$ the one that grows exponentially with time). This implies that the variance of the random variable estimated through the Monte Carlo sum grows exponentially with time, requiring an exponentially increasing number of samples to achieve a given accuracy.

Conjugating the complexified time evolution operator with the unitaries $\mathcal{U}_\mathcal{A}(\frac{\pi}{2})\mathcal{U}_\mathcal{B}(-\frac{\pi}{2})$ we find that the complexity of simulating $\Psi_{\bm n, \bm m}(t,J,U)$ and $\Psi_{\bm n, \bm m}(t,-J,U)$ is the same.

{\it Discussion.-} The method that we have presented in this work allows to approximate the time evolved wavefunction $\Psi_{\bm{nm}}(t)$ of imaginary interaction Fermi-Hubbard models for a fixed point $(\bm{n},\bm{m})$ in configuration space. There is still an exponential cost of computing all the  possible combinations of points in configuration space, just because there are combinatorially many possibilities for a fixed total spin sector. We think that setting aside this extra computational complexity in the characterisation is reasonable as even the computation of a single element is believed to be hard (i.e. not classically achievable) for general values of the parameters $(J,U)$. 
One exciting possibility towards the solution of time dynamics for real-valued Fermi-Hubbard Hamiltonians opened by our result is the use of using deformation theory and Lefschetz thimbles \cite{Mukherjee_2014,Tanizaki_2016,Fukuma_2019,Ulybyshev_2020}.  This amounts to finding a path that interpolates from the classically accessible line $({\rm arg}(J),{\rm arg}(U))=(0,\pi/2)$ to the real-valued Hamiltonian $({\rm arg}(J),{\rm arg}(U))=(0,0)$ with minimal increase in sample complexity. We have already seen that the sample complexity of our approach blows up when the model has some non-zero imaginary part of hopping amplitude $J$. This does not rule out the existence of paths in the complexity doughnut where the complexity remains bounded, or grows slowly with the evolution time. If the classical simulation of the (real valued) Fermi-Hubbard model is exponentially costly classically, as it is widely believed, then this method is bound to fail, but even understanding how it does so may reveal hidden structure in the complexity landscape.

{\it Acknowledgements.-} R. S. is pleased to thank Sabrina Y. Wang for the many discussions leading to the discovery of the algorithm for simulating Lindbladians. R.S. also thanks several of the team members within Phasecraft for the discussions that in different levels contributed to the results discussed in this work.

\bibliography{refs}

\end{document}